\documentclass{article}

\usepackage{amssymb,amsfonts,amsmath}
\usepackage{cite,enumerate,float,indentfirst}
\usepackage{color}

\def\be{\begin{eqnarray}}
\def\ee{\end{eqnarray}}
\def\nn{\nonumber}

\def\Tr{{\rm Tr}\,}

\def\a{{\bf a}}
\def\q{{\bf q}}
\def\t{{\bf t}}

\def\beq{\be }
\def\eeq{\ee}

\def\Sch{{\rm Schur}}

\def\M{{\rm Md}}

\definecolor{red}{rgb}{1,0,0}
\definecolor{orange}{rgb}{1,0.5,0}
\definecolor{violet}{rgb}{0.7,0,1}

\hoffset= - 1.0in         

\begin{document}

\title{\bf
Hopf superpolynomial from topological vertices
}

\author{
{\bf Andrei Mironov$^{a,b,c}$}\footnote{mironov@lpi.ru; mironov@itep.ru},
\ and  \  {\bf Alexei Morozov$^{d,b,c}$}\thanks{morozov@itep.ru}
}
\date{ }

\maketitle

\vspace{-5.0cm}

\begin{center}
\hfill FIAN/TD-03/20\\
\hfill IITP/TH-05/20\\
\hfill ITEP/TH-06/20\\
\hfill MIPT/TH-05/20
\end{center}

\vspace{3.cm}

\begin{center}
$^a$ {\small {\it Lebedev Physics Institute, Moscow 119991, Russia}}\\
$^b$ {\small {\it ITEP, Moscow 117218, Russia}}\\
$^c$ {\small {\it Institute for Information Transmission Problems, Moscow 127994, Russia}}\\
$^d$ {\small {\it MIPT, Dolgoprudny, 141701, Russia}}
\end{center}

\vspace{.0cm}

\begin{abstract}
Link/knot invariants are series with integer coefficients, and it is a long-standing
problem to get them positive and possessing cohomological interpretation.
Constructing positive ``superpolynomials" is not straightforward,
especially for colored invariants.
A simpler alternative is a multi-parametric generalization of the character expansion,
which leads to colored ``hyperpolynomials".
The third construction involves branes on resolved conifolds,
which gives rise to still another family of invariants
associated with composite representations.
We revisit this triality issue in the simple case of the Hopf link and
discover a previously overlooked
way to produce positive colored superpolynomials from the
DIM-governed four-point functions,
thus paving a way to a new relation between super- and hyperpolynomials.
\end{abstract}

\section{Introduction}

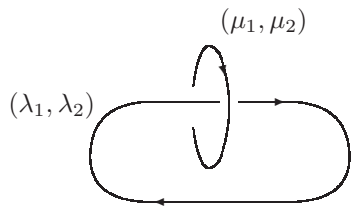
\begin{figure}[h]
\begin{picture}(100,75)(-200,-40)
\put(0,2){\line(1,0){25}}
\put(32,2){\vector(1,0){18}}

\qbezier(25,-20)(32,0)(25,20)
\qbezier(25,-20)(18,-30)(15,-8)
\qbezier(25,20)(18,30)(15,8)



\qbezier(0,2)(-24,2)(-24,-19)
\qbezier(0,-36)(-24,-36)(-24,-19)
\put(50,-36){\vector(-1,0){50}}
\qbezier(50,2)(74,2)(74,-19)
\qbezier(50,-36)(74,-36)(74,-19)

\put(-55,-2){\mbox{$(\lambda_1,\lambda_2)$}}
\put(25,30){\mbox{$(\mu_1,\mu_2)$}}
\put(25,20){\vector(1,-4){2}}

\end{picture}
\caption{{\footnotesize Hopf link
with two circles associated with two Wilson lines
in composite representations $(\lambda_1,\lambda_2)$ and $(\mu_1,\mu_2)$ accordingly, see Fig.\ref{composite}.
}}
\label{Hopfpic}
\end{figure}

For the last two decades, the duality between the two topological theories,
Chern-Simons theory \cite{CS} and topological strings,
attracts much attention and is a subject of intensive investigation.
One of the first observations of this kind was the Ooguri-Vafa correspondence \cite{OV}
that established a connection between the generating function of the Wilson averages
in Chern-Simons theory in various representations and
topological string theory on resolved conifold.
In fact, the main object in topological string theory is the topological
vertex \cite{12-13}.
It is a long-standing problem to express the knot and link invariants in these terms,
and, more generally, in terms of arbitrary tangle blocks \cite{tangles}.
The simplest  example is provided by the resolved
conifold constructed from just a pair of such vertices,
i.e. from a four-point function.
It is well known \cite{12-13} that,
for a particular choice of {\it two} non-trivial representations on external legs,
the answer coincides with that for the 2-component Hopf
link, while, for arbitrary choice of {\it four} representations on external legs,
one obtains the 2-component Hopf link colored with composite representations \cite{AKMM0},
see Fig.\ref{Hopfpic}.

The refinement of this duality that would add one more parameter
was the next important issue to address.
Adding this parameter allows one to deal with full two-parametric $\Omega$-background,
or with the Ding-Iohara-Miki algebra \cite{DIM} at generic point.
In order to preserve the duality, one needs to get a refinement of all ingredients.
The refinement of Chern-Simons theory was first proposed in \cite{ASh}
at the level Wilson averages associated with torus knots/links,
while the refined version of the Ooguri-Vafa duality was presented in \cite{Naw}.
At last, the refined topological vertex was proposed in \cite{14-17},
which opened a way to construct, at least,
the Hopf refined invariant via the refined four-point function.
Extension to more complicated knot and links requires
arbitrary convolution of vertices,
i.e. the full-fledged network models \cite{18}.

However, an unexpected problem appeared already in the Hopf case \cite{AKMM}:
while the four-point function with two non-trivial representations on the external legs
still corresponds to the refined Hopf invariant \cite{AK08,IK},
this is no longer true in the case of four non-trivial representations:
the results for the four-point function in Fig.\ref{fourpointpic}
and for the Hopf invariant in composite representations in Fig.\ref{Hopfpic}
are essentially different.
In the present letter, we find an explanation for this fact:
we argue that the refined Hopf invariant discussed earlier was the Hopf {\it hyper}polynomial
in composite representations,
while the four-point function provides a Hopf {\it super}polynomial
(which is defined to be a positive power series in appropriate variables).
These two are known to be often different in complicated representations:
already the straightforwardly constructed {\it hyper}polynomials of torus knots \cite{ASh,Che}
are not positive beyond rectangular representations, nothing to say about composite representations.
It would be a breakthrough to find an equally direct way to construct positive
expressions, which could pretend on the role of {\it super}polynomials,
and this is what we do in this paper for the Hopf link
with the help of the four-point functions.
These superpolynomials coincide with
the known Hopf hyperpolynomials in the ordinary representations $\lambda$
(which in the Hopf case are all positive),
but differ from them in the composite representations $(\lambda,\mu)$, Fig.\ref{composite},
of which the adjoint representation $([1],[1])$ is the simplest example.
Generalization to other links and knots, even torus, requires a deep insight into
topological vertex calculus and network models,
which now gets a new reason and new means for development and testing.

\section{Hyperpolynomial vs superpolynomial}

\subsection{General comments}

The standard link/knot (HOMFLY-PT) invariants depend on the two free parameters:
$q$ and $A$, and on the representations of a group.
For the group $SU(N)$, one makes a specialization $A:=q^N$.
Within the Chern-Simons realization of the invariant as a Wilson average,
the parameter $q$ is associated with $\exp\Big({2\pi i\over \kappa +N}\Big)$ where $\kappa$ is the coupling constant.
In fact, these standard invariants can be extended to include one more parameter $t$.
In this case, there are two types of invariants: hyperpolynomials and superpolynomials.
(Actually, for hyperpolynomials one can add many more parameters, see \cite{GGS,AMM,KM17fe,NawOb},
but this is beyond the scope of this letter, which concentrates on relation to superpolynomials
with a single distinguished parameter ${\bf t}=-q/t$.)

The hyperpolynomials are basically associated with torus links/knots (see, however, possible extensions in \cite{CheD,ArthSh}). There are many different but equivalent constructions of the hyperpolynomials: they can be defined either via the DAHA construction due to I. Cherednik \cite{Che}, or from the refined Chern-Simons technique of M.Aganagic and Sh.Shakirov \cite{ASh}, or by the evolution method \cite{DMMSS}. The hyperpolynomials are defined for arbitrary representations, and reduce to the standard HOMFLY-PT invariants when $t=q$, i.e. ${\bf t}=-1$.

At the same time, the notion of superpolynomial exists for all knots
but still waits for a general definition in the case of generic representations,
especially composite.
The superpolynomial was introduced in \cite{DGR} as a positive Laurent polynomial
in some specific variables that, after a proper reduction,
gives rise to the Khovanov \cite{Kh} and Khovanov-Rozhansky invariants \cite{KhR}.
At $t=q$, this polynomial reduces to the HOMFLY-PT invariant
associated with the fundamental representation of the group $SU(N)$.
However, the notion of superpolynomial has been later extended to higher representations,
symmetric or antisymmetric \cite{IGKV,AK,GS,DMMSS} and mixed \cite{AMMM,GGS,KNTZ,Mor19}.

In cases, when both the superpolynomial and the hyperpolynomial are known,
they typically coincide in symmetric and antisymmetric representations.
At the same time, they are distinct in the mixed representations:
already the trefoil hyperpolynomial contains negative terms
already in the first mixed representation $[2,1]$.
These negative terms are related to the corresponding terms in the superpolynomial
by a simple rule \cite{AMM}.

Here we consider far more complicated examples:
the hyper- and superpolynomials in the {\it composite} representations (see Fig.\ref{composite}),
but in a simpler case of the Hopf link.
There are two essentially new features in this case.
First, the superpolynomial for links
is typically not a positive polynomial (in the proper variables),
but rather a positive power series, which reduces to a finite Khovanov or Khovanov-Rozhansky polynomials after a proper reduction.
Second, the hyperpolynomial for the Hopf link appears to be positive for all
ordinary representations, including the mixed ones.
However, this is not any longer the case for the {\it composite} representations,
the simplest of which is the adjoint representation.

In this note, we present another way to construct the Hopf superpolynomials
in the composite representations: as the topological four-point function.
The so constructed polynomial is a positive series,
and it coincides with the HOMFLY-PT invariant in the composite representations
upon specialization to $t=q$.
Thus, the existence of two different continuations from this specialization
to the refined case of $t\ne q$ gets a natural explanation:
one continuation is a hyperpolynomial, the other one is a superpolynomial,
and these are not always the same.
In general the difference arises for mixed representations,
but for the Hopf link
it is postponed to more involved composite representations,
exactly the ones, which are described by the four-point functions.

\subsection{Constructing invariants}

In this section we remind the basic properties of the superpolynomials and
their relation to Khovanov-Rozhansky cohomologies.
In particular, we emphasize that generically the superpolynomial is rather a super{\it series},
and that the reduction procedures are easy to formulate and control.
We begin with the simplest and most familiar example of the trefoil knot
and then generalize it to the Hopf link.

\subsubsection{Superpolynomial}

In the fundamental representation, the unreduced trefoil superpolynomial (which coincides with the hyperpolynomial in this case) is
\be
{P}^{{\rm tref}}_{\Box}(A,q,t)\sim {\{A\}\over\{t\}}\cdot\Big(1-q^2A^2+q^2t^2\Big)
\ee
where $\{x\}:=x-1/x$. In fact, this expression becomes a polynomial only when one considers the reduced invariant: that divided by the unknot ${\{A\}\over\{t\}}$.
However, this polynomial is still not positive.
In order to get a positive polynomial, one has to make a change of variables:
$t={\bf q}$, $q=-{\bf qt}$, $A={\bf a\sqrt{-t}}$. Then,
\be\label{Ptr}
{\cal\bf P}^{{\rm tref}}_{\Box}({\bf a},{\bf q},{\bf t})\sim 1+{\bf a}^2{\bf q}^2{\bf t}^3+{\bf q}^4{\bf t}^2
\ee
This is the (reduced) positive superpolynomial in the fundamental representation for the trefoil.

\subsubsection{Reduction}

Now, in order to get the Khovanov-Rozhansky polynomials corresponding to $SU(N)$, one has first to throw away the terms
which cancel   each other at $A=t^N$, i.e. at ${\bf a}^2{\bf t}=-{\bf q}^{2N}$, and then  put $A=t^N\sqrt{q/t}$, i.e. ${\bf a=q}^N$.
Similarly, in order to get the Heegard-Floer polynomial, one has to put $A=t/q$, i.e. ${\bf a}^2{\bf t}^3=-1$ at the first step and, at the second step,  put $A=\sqrt{t/q}$, i.e. ${\bf a=t}^{-1}$. For instance, for (\ref{Ptr}) one obtains the Poincare polynomial for the Khovanov homology ($N=2$)
\be
\Pi^{{\rm tref}}_\Box(\q,\t)=1+{\bf q}^4{\bf t}^2+{\bf q}^6{\bf t}^3
\ee
since no cancellation occurs at the first step gives  in this case.

\subsubsection{Series}

A more involved is the case of links, since, in this case, there is always a denominator presented. In fact, as we already emphasized, one generally should expect the superpolynomial to be not a positive polynomial, but a positive power series. One can meet this already in the trefoil case if considering the unreduced invariant. In this case, one obtains
\be
{\cal\bf P}^{{\rm tref}}_{\Box}({\bf a},{\bf q},{\bf t})\sim{1+\a\t^2\over 1-\q^2}\Big(1+{\bf a}^2{\bf q}^2{\bf t}^3+{\bf q}^4{\bf t}^2\Big)
\ee
The denominator has to be understood as a power series. It is a positive power series
\be
{1+\a\t^2\over 1-\q^2}\Big(1+{\bf a}^2{\bf q}^2{\bf t}^3+{\bf q}^4{\bf t}^2\Big)=
\Big(1+\a\t^2\Big)\Big(1+\q^2+\q^4+\ldots\Big)\Big(1+{\bf a}^2{\bf q}^2{\bf t}^3+{\bf q}^4{\bf t}^2\Big)
\ee
Then, in order to calculate the Khovanov homology, one can note that cancelling $\a^2\t+\q^4$ gives
\be
\Big(1+\a\t^2\Big)\Big(1+\q^2+\q^4+\ldots\Big)=1+\q^2+\Big(\q^4+\a\t^2\Big)\Big(1+\q^2+\q^4+\ldots\Big)=1+\q^2
\ee
i.e. the Poincare polynomial in this case is
\be
\Pi^{{\rm tref}}_\Box(\q,\t)=\Big(1+\q^2\Big)\Big(1+{\bf q}^4{\bf t}^2+{\bf q}^6{\bf t}^3\Big)=1+\q^2+\q^4\t^2+\q^6\t^2+\q^6\t^3+\q^8\t^3
\ee

\subsubsection{Links}

Thus, the procedure works equally well for the reduced and unreduced invariants. However, for links, a denominator is always present, and, for the sake of definiteness, we deal with the unreduced invariants. Consider the simplest case of the Hopf link. In this case, \cite{IGKV,AK,DMMSS}
\be
{\cal\bf P}^{{\rm Hopf}}_{\Box}({\bf a},{\bf q},{\bf t})\sim{1+\a\t^2\over (1-\q^2)^2}
\Big(1-\q^2+{\bf a}^2{\bf q}^2{\bf t}^3+{\bf q}^4{\bf t}^2\Big)
\ee
The polynomial in the numerator is not positive. However, expanding again $1/(1-\q^2)$ into the power series, one obtains
\be
{\cal\bf P}^{{\rm Hopf}}_{\Box}({\bf a},{\bf q},{\bf t})\sim\Big(1+\a\t^2\Big)\Big(1+\q^2+\q^4+\ldots\Big)
\Big(1+\t^2\q^2({\bf a}^2{\bf t}+{\bf q}^2)\sum_{i=0}^\infty\q^{2i}\Big)
\ee
In order to calculate the Poincare polynomial for the Khovanov homology, one again notes that the first two factors after cancellation give $1+\q^2$, and similarly $({\bf a}^2{\bf t}+{\bf q}^2)\sum_{i=0}^\infty\q^{2i}$ gives after cancellation just $\q^2$. Thus, one finally obtains the Poincare polynomial
\be
\Pi^{{\rm Hopf}}_\Box(\q,\t)=(1+\q^2)(1+\q^4\t^2)=1+\q^2+\q^4\t^2+\q^6\t^2
\ee
which is, indeed, a correct and well-known answer \cite{M}.

\section{Higher representations: Hopf link hyperpolynomial}

\subsection{Superpolynomial vs hyperpolynomials}

One can develop the notion of superpolynomial for higher representations. For calculating the superpolynomials of the torus knots and links so far, it was sufficient to use their construction in terms of hyperpolynomials \cite{ASh,Che}. Note that this is quite immediate, while getting superpolynomials in a different way is a much more involved and hard problem \cite{DGR}. Still, it also turns out possible in concrete cases and for the concrete representations \cite{DMMSS,GS}. It turns out possible even to get explicit expressions for the superpolynomials of whole series of knots in (anti)symmetric representations \cite{DMMSS,IMMMfe,twist,evo} and even in mixed representations \cite{AMMM,GGS,KNTZ,Mor19}.
In all tested cases, the superpolynomials coincide with the hyperpolynomials with the exception of mixed representations: as is discussed in \cite{GGS,AMM}, the hyperpolynomials can have some negative terms in these cases. As we explain below, the hyperpolynomial for the Hopf link is positive even in mixed representations. However, it is not positive for the composite representations. We describe this in detail now.

\subsection{Hopf link hyperpolynomial}

For the Hopf link, there is a general formula for the hyperpolynomial.
The Hopf hyperpolynomial for the link components colored with the representations of $SU(N)$ associated with two Young diagrams $\lambda$ and $\mu$
is defined by the formula \cite{IK}
\be\label{Hopf}
{\cal P}_{\lambda,\mu}^{\rm Hopf} =\M_\lambda\cdot M_\mu({\bf p}_k^{*\lambda})
\ee
which coincides with the refined version \cite{DMMSS} of the Rosso-Jones formula \cite{RJ}
\be\label{RJ}
{\cal P}_{\lambda,\mu}^{\rm Hopf} =f_\lambda f_\mu
\sum_{\eta\in \lambda\otimes \mu}
{\bf N}_{\lambda\mu}^\eta f_\eta^{-1}
\cdot {\M}_\eta
\ee
Here
\be\label{fr}
f_\mu:=\left(-{q\over t}\right)^{|\mu|}q^{\nu'(\mu)}t^{-\nu(\mu)}=(-1)^{|\mu|}\ q^{\sum\mu_i^2}\ t^{-\sum{\mu^\vee_i}^2}
\ee
is the framing factor \cite{Taki} and
\be\label{nu}
\M_\eta:=M_\eta \{{\bf p}^{*\varnothing}\},\ \ \ \ \ \ \ \nu(\lambda):=2\sum_i (i-1)\lambda_i,\ \ \ \ \ \nu'(\lambda):=\nu(\lambda^\vee)
\ee
$M_\lambda(p_k)$ here is the Macdonald polynomial and
\be\label{tv}
{\bf p}^{*\mu}_k = \frac{A^k-A^{-k}}{t^k-t^{-k}}
+ A^{-k} \sum_i t^{(2i-1)k}(q^{-2k\mu_i}-1)
\ee
see \cite{AKMM} for the detailed definitions and notation.
The definition (\ref{Hopf}) coincides with \cite{ASh,DMMSS,Che}. One can check that ${\cal P}_{\lambda,\mu}^{\rm Hopf}={\cal P}_{\mu,\lambda}^{\rm Hopf}$ as it should be.
Note that in (\ref{Hopf}) we have omitted the $U(1)$-factor $q^{2|\lambda||\mu|\over N}$, \cite{Atiah,MarF,China1,tangles}.

\subsection{Positivity}

In contrast with the trefoil case, the hyperpolynomial for the Hopf link is positive in the variables $\a$, $\q$, $\t$ even in mixed representations. For instance,
\be
{\cal P}_{\Box,[2,1]}^{\rm Hopf}(\a,\q,\t)\sim{(1+\a^2\t)(\q^2+\a^2\t)(1+\a^2\q^2\t^3)\Big(1+\q^4\t^2+\q^4\t^4(\q^4+\a^2\t)
\sum_{j=0}\q^{2j}\Big)
\over (1-\q^6\t^2)(1-\q^2)^2}
\ee
This answer can be compared with the trefoil case \cite[sec.4.5]{AMM}, where the hyperpolynomial is not positive even in the first mixed representation $[2,1]$. A more involves answer for the case of the both link components colored with the mixed representation $[2,1]$ can be found in the Appendix.

\begin{figure}
\begin{picture}(300,125)(-90,-30)

\put(0,0){\line(0,1){90}}
\put(0,0){\line(1,0){250}}
\put(50,40){\line(1,0){172}}

\put(0,90){\line(1,0){10}}
\put(10,90){\line(0,-1){20}}
\put(10,70){\line(1,0){20}}
\put(30,70){\line(0,-1){10}}
\put(30,60){\line(1,0){10}}
\put(40,60){\line(0,-1){10}}
\put(40,50){\line(1,0){10}}
\put(50,50){\line(0,-1){10}}

\put(265,2){\mbox{$\vdots$}}
\put(265,15){\mbox{$\vdots$}}
\put(265,28){\mbox{$\vdots$}}

\put(252,0){\mbox{$\ldots$}}
\put(253,40){\mbox{$\ldots$}}
\put(239,40){\mbox{$\ldots$}}
\put(225,40){\mbox{$\ldots$}}

\put(222,40){\line(0,-1){10}}
\put(222,30){\line(1,0){10}}
\put(232,30){\line(0,-1){20}}
\put(232,10){\line(1,0){18}}
\put(250,0){\line(0,1){10}}

\put(0,90){\line(1,0){10}}
\put(10,90){\line(0,-1){20}}
\put(10,70){\line(1,0){20}}
\put(30,70){\line(0,-1){10}}
\put(30,60){\line(1,0){10}}
\put(40,60){\line(0,-1){10}}
\put(40,50){\line(1,0){10}}
\put(50,50){\line(0,-1){10}}

{\footnotesize
\put(123,17){\mbox{$ \overline{\mu}$}}
\put(17,50){\mbox{$\lambda$}}
\put(243,22){\mbox{$\check \mu$}}
\qbezier(270,3)(280,20)(270,37)
\put(280,18){\mbox{$h_\mu = l_{\mu^{\vee}}=\mu_{_1}$}}
\qbezier(5,-5)(132,-20)(260,-5)
\put(130,-25){\mbox{$N $}}
\qbezier(5,35)(25,25)(45,35)
\put(22,20){\mbox{$l_\lambda$}}
\qbezier(225,43)(245,52)(265,43)
\put(243,52){\mbox{$l_{\!_\mu}$}}
}

\put(4,40){\mbox{$\ldots$}}
\put(18,40){\mbox{$\ldots$}}
\put(32,40){\mbox{$\ldots$}}

\end{picture}
\caption{{\footnotesize
Composite are representation of $SU(N)$ described by the $N$-dependent
Young diagram
$(\lambda,\mu)= \Big[\lambda_1+\mu_1,\ldots,\lambda_{l_R}+\mu_1,\mu_1,\ldots,\mu_1,\mu_1-\mu_{_{l_{\!_\mu}},
\mu_1-\mu_{_{l_{\!_\mu}-1}}},\ldots,\mu_1-\mu_2\Big]$ with $\mu_1$ repeated $N-l_{\!_\lambda}-l_{\!_\mu}$ times.
The ordinary $N$-independent representations in this notation are $\lambda=(\lambda,\varnothing)$,
there conjugate are $\overline{\lambda} = (\varnothing,\lambda)$.
The simplest of non-trivial composite representations is the
adjoint $(1,1) = [2,1^{N-2}]$.
}}
\label{composite}
\end{figure}
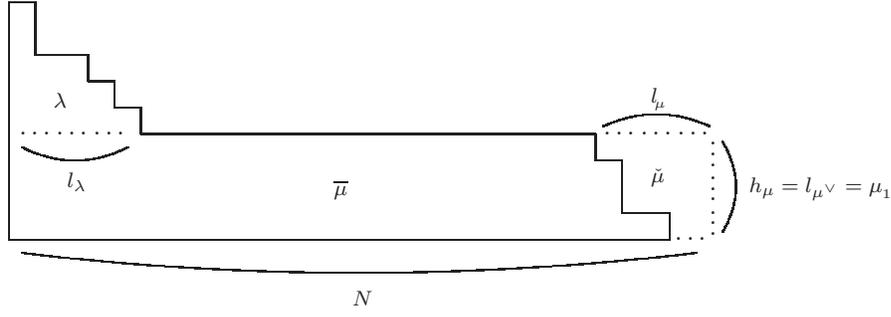

We made a computer check for all the representations with $|\lambda|\le 8$, $|\mu|\le 8$ that the first tens of terms in the power series expansion of (\ref{Hopf}) are positive in the boldface variables $\a$, $\q$, $\t$. Hence, one may expect that the hyperpolynomial in the Hopf case coincides with the superpolynomial.
However, the situation changes as soon as one considers more general class of representations:
the composite representations \cite{comprep,MMhopf}, of which the adjoint representation is the simplest example. The main difference with the previous examples is that now one considers the invariant not at the fixed Young diagram, independent of the size of group, $N$, but changes the diagram along with the group.
For instance, the adjoint representation is given by the Young diagram $[2,1^{N-1}]$. Generally, the composite representation is described by a pair of Young diagrams $\lambda$ and $\mu$,
see Fig.\ref{composite},
we denote it $(\lambda,\mu)$
and refer the reader to \cite[sec.3]{AKMM} for a technical description.
Adjoint representation in these terms is $([1],[1])$.

It turns out that, for the composite representations,  there is a uniformization that allows one to express the knot/link invariant as a polynomial or a power series of $q$, $t$ and $A=t^N$ \cite{MMuniv,tangles,MMhopf,AKMM,MM}.
However, one can not expect that this invariant has a positivity property,
and, indeed, it does not.
For instance, the Hopf hyperpolynomial with one component being colored by the fundamental representation, and the other one, by the adjoint is given by the series (see \cite[eq.(17)]{AKMM} and (\ref{H1}) below)
\be
{\cal P}_{\Box,([1],[1])}^{\rm Hopf}(\a,\q,\t)=\q^2+2\q^4+5\t\q^4\a^2+3\t\q^2\a^2+3\q^4\t^3\a^2+\t\a^2-\t^5\q^4\a^2+\nn\\
+2\t^4\q^2\a^4+8\t^4\q^4\a^4+\t^2\a^4+2\t^2\q^2\a^4+3\q^4\t^2\a^4+2\q^2\t^5\a^6+5\q^4\t^5\a^6+\ldots
\ee
Similarly, in the case of the both component colored with the adjoint,
one gets (see \cite[eq.(20)]{AKMM} and (\ref{H2}) below)
\be
{\cal P}_{([1],[1]),([1],[1])}^{\rm Hopf}(\a,\q,\t)=\q^2+2\q^4+3\t\q^2\a^2+\t\a^2-\t^5\q^4\a^2+5\t\q^4\a^2+7\q^4\t^3\a^2
+\q^2\t^3\a^2+3\q^4\t^2\a^4+\nn\\
+2\t^2\q^2\a^4
+\t^4\a^4+6\t^4\q^2\a^4+\t^2\a^4+18\t^4\q^4\a^4+11\q^4\t^5\a^6+5\q^2\t^5\a^6+\t^5\a^6+\ldots
\ee
Note that the procedure of reduction to the finite $N$ as described above is not applicable in this case, since the parameter $A$ now plays an absolutely different role.

The lack of positivity may not look surprising, because no Khovanov-Rozhansky cohomology
is known for composite representations.
However, it turns out that one can still construct a counterpart of the hyperpolynomial
that has the right HOMFLY-PT limit,  and possesses the positivity property.
We explain this in the next section.

\section{Topological vertex and 4-point function}

\subsection{Unrefined case}

\begin{figure}
\begin{picture}(200,100)(-200,-50)
\put(0,0){\line(1,1){30}}
\put(30,30){\line(1,0){30}}
\put(30,30){\line(0,1){30}}
\put(0,0){\line(-1,0){30}}
\put(0,0){\line(0,-1){30}}
\put(10,10){\line(-1,-1){2}}
\put(40,30){\line(-1,0){2}}
\put(30,40){\line(0,-1){2}}
\put(-10,0){\line(1,0){2}}
\put(0,-10){\line(0,1){2}}
\put(10,20){\mbox{$\xi$}}
\put(55,35){\mbox{$\lambda_1$}}
\put(15,55){\mbox{$\mu_1$}}
\put(-28,5){\mbox{$\lambda_2$}}
\put(5,-28){\mbox{$\mu_2$}}
\end{picture}
\caption{{\footnotesize
The simplest network model, the four-point function
made from contraction of just two topological vertices
is associated with the Hopf link HOMFLY-PT invariant
in two composite representations $(\lambda_1,\lambda_2)$ and $(\mu_1,\mu_2)$.
Its refined version does not coincide with the Hopf link {\it hyper}polynomial \cite{AKMM},
but can instead provide a {\it positive} expression,
which we suggest to interpret as a {\it super}polynomial for the
Hopf link in composite representations.
}}
\label{fourpointpic}
\end{figure}
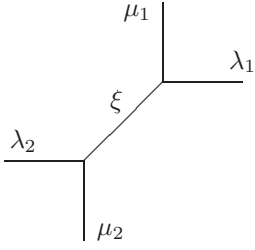

In the unrefined case $t=q$, there is an alternative way to construct the Hopf link invariant:
via the topological four-point function. That is,
one can construct the four-point function, Fig.\ref{fourpointpic},

\bigskip

\be\label{Hopfcr}
Z_{\mu_1,\mu_2;\lambda_1,\lambda_2} = \sum_\xi (-A^2)^{|\xi|} C_{\xi\mu_1\lambda_1} C_{\xi^\vee\mu_2\lambda_2}
\ee

\noindent
by the convolution of two topological vertices,
\be
C_{\xi\mu\lambda} =
q^{\nu'(\lambda)-\nu(\lambda)}\cdot \Sch_{\mu}\{p^{(\varnothing)}\}\cdot
\sum_{\eta} \Sch_{\xi/\eta} \{p^{(\mu)}\}\cdot  \Sch_{\lambda^\vee/\eta} \{p^{(\mu^\vee)}\}
\ee
where  $\Sch_{\mu}$ denotes the Schur function and the time variables $p^{(\mu)}_k$ are defined as
\be\label{pmu}
p_k^{(\mu)}  = \sum_{i=1}^\infty  q^{2\mu_ik}t^{(1-2i)k} =
\frac{1}{t^{k}-t^{-k}} + \sum_{i=1}^\infty t^{(1-2i)k}(q^{2\mu_ik}-1)=
\frac{1}{t^{k}-t^{-k}}+(q^k-q^{-k})\sum_{i,j\in\mu} t^{(1-2i)k}q^{(2j-1)k}
\ee
at $t=q$.

It turns out \cite{AKMM0} that this four-point function
is proportional to the Hopf link HOMFLY-PT invariant colored with two composite representations,
${\cal H}^{\rm Hopf}_{(\mu_1,\mu_2),(\lambda_1,\lambda_2)}$,
\be\label{eq}
\frac{Z_{\mu_1,\mu_2;\lambda_1,\lambda_2}}
{Z_{\varnothing,\varnothing:\varnothing, \varnothing}}
={\cal H}^{\rm Hopf}_{(\lambda_1,\lambda_2),(\mu_1,\mu_2)}
\ee
However, the deformation to $t\ne q$ destroys this identity: the r.h.s. made from the topological four-point function is no longer equal to the Hopf hyperpolynomial (\ref{Hopf}), \cite{AKMM}. However, it turns out that the refined four-point function allows one (after a proper changing of the basis) to obtain a positive series deformation! This is our main point.

\subsection{Refined four-point functions}

When $t\ne q$, there are a few new phenomena. First of all, there are three different four-point functions \cite[eqs.(65)-(67)]{AKMM}: two of them associated with the internal edge along the unpreferred directions are related by a simple transformation \cite[eq.(85)]{AKMM}, the third one corresponds to the internal edge along the preferred direction. The first four-point function is constructed from the refined topological vertex
\beq\label{rtv}
C_{\lambda}\!\phantom{.}^{\mu\xi} (q,t)  =
f^{-1}_\xi\cdot M_{\mu}\{p^{(\varnothing)}\}
\sum_{\eta} \Big({q\over t}\Big)^{|\eta|-|\xi|}M_{\xi/\eta}\Big\{p_k^{(\mu)}\Big\}\cdot  \overline{M}_{\lambda^\vee/\eta^\vee} \Big\{(-1)^{k+1}\overline{p}_k^{(\mu^\vee)}\Big\}
\eeq
as
\beq\label{4p}
{\bf Z}\left[\begin{array}{c|c}\!\! \mu_1& \lambda_2\\
\lambda_1& \mu_2\end{array}\right]
 &:=& \sum_{\xi} (-Q)^{|\xi|} C_{\lambda_1}^{~~\mu_1\xi}(q,t)
C_{\lambda_2}^{~~~ \mu_2\xi^\vee}(t,q)
\eeq
where $Q=A^2q/t$. Here bar denotes the function with $q$ and $t$ interchanged. Note that, in the refined case, the lower and upper indices in the vertex are different, in contrast with the unrefined case, where there is a cyclic symmetry of all three indices.

\bigskip

This expression with the normalization factor as in (\ref{eq}) can be rewritten as \cite{AK08}
\be
\widehat{\bf Z}\left[\begin{array}{c|c}\!\! \mu_1& \lambda_2\\
\lambda_1& \mu_2\end{array}\right]:=
{\bf Z}\left[\begin{array}{c|c}\!\! \mu_1& \lambda_2\\
\lambda_1& \mu_2\end{array}\right]\cdot
{\bf Z}\left[\begin{array}{c|c}\!\! \varnothing& \varnothing\\
\varnothing& \varnothing\end{array}\right]^{-1}
=
(-1)^{|\lambda_1|+|\lambda_2|}A^{|\mu_1|+|\mu_2|+|\lambda_1|+|\lambda_2|}\left({q\over t}\right)^{|\mu_2|+|\lambda_1|}f_{\mu_1}\overline{f}_{\mu_2}
D_{(\mu_1,\mu_2)}\times
\nn
\ee
\vspace{-0.4cm}
\be
\times
\sum_\sigma (-1)^{|\sigma|}\cdot\overline{M}_{\lambda_1^\vee/\sigma}\Big\{\Big({Aq\over t}\Big)^k\overline{p}^{(\mu_2)}_k-\Big({Aq\over t}\Big)^{-k}\overline{p}^{(\mu_1^\vee)}_k
\Big\}\cdot
M_{\lambda_2^\vee/\sigma^\vee}\Big\{A^{k}p^{(\mu_1)}_k-A^{-k}p^{(\mu_2^\vee)}_k\Big\}
\label{4p1}
\ee
where
\be
D_{(\mu_1,\mu_2)}:=\sum_\sigma (-1)^{|\sigma|}
M_{\mu_1/\sigma} \{{\bf p}^{*\varnothing}\}\Big|_{A\to Aq/t}M_{\mu_2/\sigma^\vee} \{{\bf p}^{*\varnothing}\}
\ee
and $M_{\lambda/\mu}$ is the skew Macdonald polynomial \cite{Mac}.

\bigskip

There are two more possibilities: one can convert topological vertices with internal edge along the other unpreferred direction
\beq\label{Z2}
{\bf Z'}\left[\begin{array}{c|c}\mu_1\,\lambda_1&\!\! \\\!\! &\mu_2\,\lambda_2\end{array}\right]
&:=& \sum_{\xi} (-Q)^{|\xi|} C_{\xi }^ {~~\mu_1 \lambda_1}(q,t)
C_{\xi^\vee}^{~~~ \mu_2\lambda_2}(t,q)
\eeq
or along the preferred direction
\beq\label{Z3}
{\bf Z''}\left[\begin{array}{c|c}\!\! \lambda_3 & \lambda_4 \\
\lambda_1& \lambda_2\end{array}\right]&:=& \sum_{\xi} (-Q)^{|\xi|} C_{\lambda_1}^{~~ \xi \lambda_3}(q,t)
C_{\lambda_4}^{~~ \xi^\vee \lambda_2}(t,q)
\eeq
In the first case, the topological four-point function $\widehat{\bf Z'}$
with the internal edge along the unpreferred direction, \cite[eqs.(66) and (84)]{AKMM}
is related to (\ref{4p1}), \cite[eq.(85)]{AKMM}: it is described by some interchanging and conjugating the external Young diagram and by the change of variables $\a\to\a^{-1}$, $\t\to\t^{-1}$, $\q\to -\q\t$:
\be\label{flop}
\widehat{\bf Z'}\left[\begin{array}{c|c}\mu_1\,\lambda_1&\!\! \\\!\! &\mu_2\,\lambda_2\end{array}\right](\a,\q,\t) \ \ \sim \ \
\widehat{\bf Z}\left[\begin{array}{c|c}\!\! \mu_2& \lambda_2^\vee\\
\lambda_1^\vee& \mu_1\end{array}\right](\a^{-1},-\q\t,\t^{-1})
\ee
The third four-point function may look more involved, since it has no a representation in the form of a simple finite sum like (\ref{4p1}). However, it is known \cite{AK} that the normalized four-point function is a polynomial in $A^2$ of degree $|\mu_1|+|\mu_2|+|\lambda_1|+|\lambda_2|$. Hence, it is enough to expand this expression into the Taylor series up to this order in $A$ making the calculations of almost the same level of difficulties as the other four-point functions.

\subsection{Changing the basis and relation with the Hopf link}

Note that the four-point function actually depends in the refined case on the choice of the basis. Indeed, an invariant is the open string partition function with four holonomies $U_i$, $i=1,2,3,4$,
\be
Z_{str}(U_1,U_2,U_3,U_4):=\sum_{\mu_i,\lambda_i}
\widehat{\bf Z}\left[\begin{array}{c|c}\!\! \mu_1& \lambda_2\\
\lambda_1& \mu_2\end{array}\right]\cdot V_{\mu_1}(U_1)
V_{\mu_2}(U_2) V_{\lambda_1}(U_3)V_{\lambda_2}(U_4)
\ee
where $V_\mu(U)$ is a (graded) basis of symmetric functions. In the unrefined case, there is no much choice: one would naturally choose as $V_\mu(U)$ the Schur functions. On the contrary, in the refined case, there are a few natural choices: $M_{\mu}(U)$, $\overline{M}_{\mu}(U)$, etc., \cite{AKMM}. In \cite{IK}, the authors considered the case when only two of the external lines in the four-point function carry out a non-trivial Young diagram and chose $V_\mu(U)=\overline{M}_{\mu}(-\xi_k)$, $\xi_k:=\Tr U^k$ for the external $\mu$ corresponding to the lower index in the topological vertex (\ref{rtv}) in order to compared
this partition function with the generating function of Hopf link invariants in the refined Chern-Simons theory, where $V_\mu(U)$ are $\overline{M}_{\mu}(U)$.

In fact, all the three topological four-point functions almost in all cases of two external Young diagrams trivial reduce either to the Hopf hyperpolynomial, or to the Macdonald dimension. It is sometimes an immediate connection, in other cases it requires this just described changing of basis (see \cite[sec.8]{AKMM}). In fact, in accordance with \cite[sec.8]{AKMM}, in order to get the usual Hopf hyperpolynomial ${\cal P}_{\lambda,\mu}^{\rm Hopf}$, one has to change in the four-point function (\ref{4p}) the basis $\overline{M}_{\lambda}(\xi_k) \to\overline{M}_{\lambda}(-\xi_k)$, i.e. effectively to remove an additional minus sign of the argument of the rightmost  Macdonald polynomial in (\ref{rtv}).
In other words, this changing the basis is defined by the relation
\be\label{cb1}
\sum_{\mu_i,\lambda_i}
\widehat{\bf Z}\left[\begin{array}{c|c}\!\! \mu_1& \lambda_2\\
\lambda_1& \mu_2\end{array}\right]\cdot \overline{M}_{\mu_1}(\xi^{(1)}_k)\overline{M}_{\mu_2}(\xi^{(2)}_k){M}_{\lambda_1}(-\xi^{(3)}_k)
{M}_{\lambda_2}(-\xi^{(4)}_k)=\nn\\
=\sum_{\mu_i,\lambda_i}
\widehat{\bf Z}_{cb}\left[\begin{array}{c|c}\!\! \mu_1& \lambda_2\\
\lambda_1& \mu_2\end{array}\right]\cdot \overline{M}_{\mu_1}(\xi^{(1)}_k)\overline{M}_{\mu_2}(\xi^{(2)}_k){M}_{\lambda_1}(\xi^{(3)}_k)
{M}_{\lambda_2}(\xi^{(4)}_k)
\ee
which is equivalent to changing the basis for {\it some} of the topological vertices.
We label the modified ${\bf Z}$-function by index $cb$ (changed basis).
Similarly, one can make  the basis change in the case of the four-point function
with internal edge along the preferred direction in such a way that all four
signs on the external legs are changed:
\be\label{cb2}
\sum_{\lambda_i}
\widehat{\bf Z}''\left[\begin{array}{c|c}\!\! \lambda_3& \lambda_4\\
\lambda_1& \lambda_2\end{array}\right]\cdot {M}_{\lambda_1}(-\xi^{(1)}_k)\overline{M}_{\lambda_2}(-\xi^{(2)}_k)\overline{M}_{\lambda_3}(-\xi^{(3)}_k)
{M}_{\lambda_4}(-\xi^{(4)}_k)=\nn\\
=\sum_{\lambda_i}
\widehat{\bf Z}_{cb}''\left[\begin{array}{c|c}\!\! \lambda_3& \lambda_4\\
\lambda_1& \lambda_2\end{array}\right]\cdot {M}_{\lambda_1}(\xi^{(1)}_k)\overline{M}_{\lambda_2}(\xi^{(2)}_k)\overline{M}_{\lambda_3}(\xi^{(3)}_k)
{M}_{\lambda_4}(\xi^{(4)}_k)
\ee
In fact, there are many more possibilities even only in playing with signs: one can change the signs in the arguments of any set of these four Macdonald polynomials.

Our main claim here is that {\bf the four-point functions like  (\ref{4p}) or (\ref{Z3}) lead to positive series in the $\a$, $\q$, $\t$ variables, sometimes in a basis, modified like in (\ref{cb1}) and (\ref{cb2}).}

\subsection{Adjoint representation}

Let us consider the two simplest examples, when one of the Hopf component is colored with the adjoint representation, and the one, with either the fundamental representation, or with the adjoint one. In this case, there is no problem of choosing the basis, since there is only one function at the first level. Moreover, in this case, all the three topological four-point functions coincide. One can borrow the answers both for the Hopf hyperpolynomial and for the four-point function in these cases from \cite{AKMM}, they look like

{\footnotesize
\be\label{H1}
{\cal P}_{\Box,([1],[1])}^{\rm Hopf}={\{Aq\}\{A\}\{A/t\}\over \{t\}^3\{Aq/t\}}\cdot\left[\Big(q^2-(q/t)^2+t^{-2}\Big)A-\Big(t^2-(t/q)^2+q^{-2}\Big)A^{-1}\right]\sim
\nn\\ \nn \\
\sim {(1+\a^2\q^2\t^3)(1+\a^2\t)(\q^2+\a^2\t)\over (1+\a^2\t^3)(1-\q^2)^2}
\Big(1-\a^2\q^2\t^5+\q^4\t^2\sum_{i=0}^\infty \q^{2i}\Big)\sim \q^2+2\q^4+(\t+3\q^2t+5\q^4\t+3\q^4\t^3-\q^4\t^5)\a^2+\ldots
\nn
\ee
}

\noindent
which is not positive, while

{\footnotesize
\be\label{18}
{\cal\bf P}_{\Box,([1],[1])}^{\rm Hopf}:=\widehat{\bf Z}\left[\begin{array}{c|c}\!\! [1]& \varnothing\cr
[1]& [1]\end{array}\right]\sim {\{Aq\}\{A/t\}\{Aq/t\}\over \{t\}\{q\}^2}\Big(t^2-(t/q)^2+q^{-2}\Big)
\sim {(1+\a^2\q^2\t^3)(1+\a^2\t)(\q^2+\a^2\t)\over (1+\a^2\t^3)(1-\q^2\t^2)^2}
\Big(1+\q^4\t^2\sum_{i=0}^\infty \q^{2i}\Big)\nn
\ee
}

\noindent
is clearly positive.

\bigskip

Similarly, in the second case,

{\footnotesize
\be\label{H2}
{\cal P}_{([1],[1]),([1],[1])}^{\rm Hopf}={\{Aq\}\{A/t\}\{A\}^2\{q\}\over \{t\}^3\{Aq/t\}^2}\left(-1+{\{Aq/t\}\over
\{A\}\{q\}\{t\}}\left[A\cdot\Big(q^2-(q/t)^2+t^{-2}\Big)
-\frac{1}{A}\cdot \Big(t^2-(t/q)^2+q^{-2}\Big)\right]^2\right)=\nn\\ \nn \\
={(1+\a^2\q^2\t^3)(1+\a^2\t)^2(\q^2+\a^2\t)(1-\q^2\t^2)\over \a^4\q^2\t^6(1+\a^2\t^3)^2(1-\q^2)^3}
 \left(\a^2\q^2\t^5+{1+\a^2\t^3\over 1+\a^2\t}\left[\sum_{i=1}^\infty \q^{2i}+\a^2\t^3\sum_{i=1}^\infty (\q\t)^{2i}-
{1+\a^2\t^3\over (1-\q^2)(1-\q^2\t^2)}\right]^2
\right)\sim\nn\\ \nn \\
\sim \q^2+2\q^4+3\t\q^2\a^2+\t\a^2-\t^5\q^4\a^2+5\t\q^4\a^2+7\q^4\t^3\a^2
+\q^2\t^3\a^2+\ldots
\nn
\ee
}

\noindent
which contains negative terms, while the four-point function is

{\footnotesize
\be
{\cal\bf P}_{([1],[1]),([1],[1])}^{\rm Hopf}=\widehat{\bf Z}\left[\begin{array}{c|c}\!\! [1]&[1]\cr
[1]&[1]\end{array}\right]\sim
{\{Aq\}\{A/t\}\over \{t\}\{q\}}\left(-1+{\{A\}\{Aq/t\}\over \{q\}\{t\}} \Big(q^{-2}-(q/t)^2+t^2\Big)\cdot\Big(q^2-(t/q)^2+t^{-2}\Big)\right)\sim\nn\\ \nn \\
\sim {(1+\a^2\q^2\t^3)(\q^2+\a^2\t)\over (1-\q^2)(1-\q^2\t^2)}
\left(\a^2\q^2\t^3+(1+\a^2\t)(1+\a^2\t^3)\Big(1+\q^4\t^2\sum_{i=0}^\infty \q^{2i}\Big)\Big(1+\q^4\t^2\sum_{i=0}^\infty (\q\t)^{2i}\Big)\right)
\nn
\ee
}

\noindent
which is also evidently a positive power series.

\subsection{Generic case}

In the case of higher representations, one often needs changing the basis described above in order to get the positive power series out of the topological 4-point function. This changing the basis means one considers instead of (\ref{4p1}) the four-point function
\be\label{main}
\begin{array}{c}
\widehat{\bf Z}_{cb}\left[\begin{array}{c|c}\!\! \mu_1& \lambda_2\\
\lambda_1& \mu_2\end{array}\right]
=
(-1)^{|\lambda_1|+|\lambda_2|}A^{|\mu_1|+|\mu_2|+|\lambda_1|+|\lambda_2|}\left({q\over t}\right)^{|\mu_2|+|\lambda_1|}f_{\mu_1}\overline{f}_{\mu_2}
D_{(\mu_1,\mu_2)}\times
\cr
\cr
\times
\sum_\sigma (-1)^{|\sigma|}\cdot\overline{M}_{\lambda_1^\vee/\sigma}\Big\{-\Big({Aq\over t}\Big)^k\overline{p}^{(\mu_2)}_k+\Big({Aq\over t}\Big)^{-k}\overline{p}^{(\mu_1^\vee)}_k
\Big\}\cdot
M_{\lambda_2^\vee/\sigma^\vee}\{-A^{k}p^{(\mu_1)}_k+A^{-k}p^{(\mu_2^\vee)}_k\}
\end{array}
\nn\\
\ee
We have checked for many various representations with the computer that the expansion of this function often does not contain negative terms.

When it is negative, it is often that the four-point function $Z''$ with the basis changed as in (\ref{cb2}) is positive. In other cases, positive is $Z''$ without changing the basis, (\ref{Z3}).
For instance, of 256 possible combinations of four Young diagrams $\mu_{1,2}$, $\lambda_{1,2}$ up to level 2, 148 are associated with positive four-point function (\ref{main}). Of 108 remaining possibilities, 32 are associated with positive $Z''_{cb}$ in the changed basis, (\ref{cb2}), etc.

Thus, we propose {\bf a conjecture}:

\bigskip

\fbox{\begin{minipage}{0.9\linewidth}{For any quadruple of Young diagrams, one can always choose a proper topological four-point function in variables $\a$, $\q$, $\t$ treated as a power series in $\q$ such that it is positive and gives a Hopf superpolynomial in the composite representations given by these Young diagrams.}\end{minipage}}

\bigskip

We checked this claim with all the diagrams up to level 2 and for numerous selected quadruples of bigger sizes.  Note that the answers for all the diagrams of level two are quite non-trivial (see the Appendix for a simpler example of two diagrams of level one and two diagrams of level two when the result is already quite involved), and the freedom of choosing the basis is far not enough to guarantee their positivity, it comes as a very non-trivial statement.
Of course, this is not sufficient for making a conclusive statement, but clearly means something that should be analyzed by more powerful methods than random checks. One may expect that generally one will need a more complicated change of basis, not just changing signs in the Macdonald polynomial argument as in (\ref{cb1}) and (\ref{cb2}). This is why we started with the example of the adjoint representation, when there is no freedom in choosing the basis and all the four-point functions coincide.

\section{Conclusion}

To conclude, we found a way to get a positive superpolynomial for the Hopf link
in two arbitrary composite representations by adjusting the basis in the four-point function,
made from a pair of refined topological vertices.
At the moment, the evidence in favour of this possibility
is provided by computer experiment and is limited to
rather simple representations and/or to the lowest (but quite a few) terms in the series.
These are tedious calculations and a new look is needed to convert this observation into
a well-established and well-grounded fact.
Even more interesting would be an expression for generic colored knot polynomials
(beyond the Hopf link) in terms of network models,
which would allow one to make and test more general positivity conjectures.

\section*{Acknowledgements}

This work was supported by the Russian Science Foundation (Grant No.16-12-10344).


\section*{Appendix}

This Appendix contains longer formulas, which we use to illustrate three important phenomena.

\bigskip

{\bf The first example} demonstrates that, in variance with the simple knots, the
Hopf link is {\it positive} in the mixed representations on the both components,
of which the simplest is $[2,1]$:



{\footnotesize
\be
{\cal P}_{[2,1],[2,1]}^{\rm Hopf}(\a,\q,\t)\sim {(1+\a^2\t)(\q^2+\a^2\t)(1+\a^2\q^2\t^3)\over (1-\q^2)(1-\q^6\t^2)}\times\nn\\
\times\left(\q^{10}\t^{10}\Big(\q^{10}+\a^2\q^4\t(1+\q^2+\q^6\t^2)+\a^4\q^2\t^2(1+\q^2\t^2+\q^4\t^2)
+\a^6\t^5\Big)\sum_{i=0}\q^{6i}\t^{2i}\Big(\sum_{j=0}\q^{2j}\Big)^3+\right.\nn\\  \nn \\
+\q^4\t^6\Big(\q^6(1+2\q^6(1+\t^2))+\a^2\q^2\t(1+\q^4\t^2)(1+3\q^2+\q^6\t^2)
+\a^4\t^2(\q^4\t^2+1)(\q^4\t^2+\q^2\t^2+1)\Big)\sum_{i=0}\q^{6i}\t^{2i}\Big(\sum_{j=0}\q^{2j}\Big)^2+\nn\\
\left.+\Big(1+2\q^8\t^4(1+\q^6\t^2)+3\q^{12}\t^6+\a^2\q^4\t^5(2+\q^2+\q^4\t^2)\Big)\sum_{i=0}\q^{6i}\t^{2i}\sum_{j=0}\q^{2j}
+\q^4\t^2\sum_{i=0}\q^{6i}+\t^2(\q^4+\a^2\t)\Big(\sum_{j=0}\q^{2j}\Big)^2
\right)\nn
\ee
}

\noindent

{\bf The second example} demonstrates the positivity
of the Hopf superpolynomial when one component is colored
with the adjoint representation and the other one with the composite representation $([2],[2])$.
The answer is given by formula (\ref{main})

{\footnotesize
\be
\!\!\!\!\!\!\!\!\!
\widehat{\bf Z}'\left[\begin{array}{c|c}
[2]&[1]\cr
[1]&[2]\end{array}\right]\sim{(1+\a^2\t)(1+\a^2\t^3)(\q^2+\a^2\t)(1+\a^2\q^6\t^5)\over (1-\q^4\t^2)^2}
\left(\q^8\t^4\Big[1+\q^2+\q^6\t^2\sum_{i=0}\q^{2i}\Big]\Big[1+\q^2\t^2+\q^6\t^4\sum_{i=0}(\q\t)^{2i}\Big]
\sum_{i,j=0}\q^{2(i+j)}\t^{2j}+\right.  \nn\\
+\a^2\q^2\Big[(1+\q^2\t^2)^2(1+\q^2)^2+\q^4\t^2(1+\q^2)(1+\q^2\t^2)(3+\q^4\t^2)
\sum_{i,j=0}\q^{2(i+j)}\t^{2j}+
\q^{12}\t^6(1+\q^2\t^4)\sum_{i=0}\q^{2i}\Big(\sum_{j=0}(\q\t)^{2j}\Big)^2+
\ \ \ \ \ \ \ \ \ \ \ \ \ \   \nn\\
+\q^8\t^4(1+\q^2\t^2+\q^4\t^4+\q^4\t^2+\q^6)\sum_{i=0}(\q\t)^{2i}\Big(\sum_{j=0}\q^{2j}\Big)^2+
\q^{14}\t^8(\q^2+\t^2)\Big(\sum_{i,j=0}\q^{2(i+j)}\t^{2j}\Big)^2\Big]+\nn
\ \ \ \ \ \  \ \ \ \ \ \ \ \ \ \ \ \ \ \ \ \ \ \ \ \ \
\ee
\vspace{-0.4cm}
\be
\left.+\a^4\t\Big[1+\q^2+\q^4\t^2(1+\q^2)\sum_{i=0}\q^{2i}+\q^8\t^4\Big(\sum_{j=0}\q^{2j}\Big)^2\Big]
\Big[1+\q^2\t^2+\q^4\t^2(1+\q^2\t^2)\sum_{i=0}(\q\t)^{2i}+\q^8\t^4\Big(\sum_{j=0}(\q\t)^{2j}\Big)^2\Big]
\right)
\label{HH1}
\ee}

\bigskip

{\bf The third example} shows that our suggestion about Hopf {\it super}polynomials
does not respect the obvious symmetries of the problem,
i.e. the superpolynomial is not unique.
Namely, there is no symmetry w.r.t. permutation $(\mu_1,\mu_2)\leftrightarrow(\lambda_1,\lambda_2)$, which corresponds to permuting the Hopf components, but only w.r.t. $(\mu_1,\lambda_1)\leftrightarrow(\mu_2,\lambda_2)$  \cite{AKMM}.
This is reflected in the fact that positive is only ${\footnotesize \widehat{\bf Z}_{cb}\left[\begin{array}{c|c}
[2]&[1]\cr
[1]&[2]\end{array}\right]}$, while, for the permuted representations, one has to choose another four-point function in order to get a positive superpolynomial. For instance, positive is the series,
which is different from (\ref{HH1}):

{\footnotesize
\be\label{HH2}
\!\!\!\!\!\!
\widehat{\bf {Z}}''\left[\begin{array}{c|c}
[1]&[2]\cr
[2]&[1]\end{array}\right]\sim{(1+\a^2\t)(1+\a^2\t^3)(\q^2+\a^2\t)\over (1-\q^4\t^2)^2}
\left(\Big[1+\q^2+\q^2\t+\q^6\t^4\sum_{i=0}(\q\t)^{2i}\Big]\Big[1+\q^2\sum_{i=0}(\q\t)^{2i}+\q^6\t^2
\sum_{i,j=0}\q^{2(i+j)}\t^{2j}\Big]\sum_{i=0}\q^{2i}+\right.
\ \ \ \ \ \ \ \ \ \nn\\
\!\!\!\!\!\!\!\!\!\!\!\!
+\a^2\q^2\t^3\Big[1+\q^8\t^4+\q^2\t^2(1+\q^2)(1+\q^4\t^2)\sum_{i=0}\q^{2i}+
\q^2(1+\q^4\t^2)\Big(\sum_{i=0}(\q\t)^{2i}\Big)^2+\Big(1+\q^2(1+\q^8\t^4)(1+\t^2)\Big)
\sum_{i,j=0}\q^{2(i+j)}\t^{2i}+ \ \ \ \
\nn\\
\!\!\!\!\!\!\!\!\!\!\!\!\!\!\!\!\!\!\!\!\!\!\!\!
+\q^6\t^2(1+2\q^4\t^2+\q^2\t^2)\Big(\sum_{i=0}(\q\t)^{2i}\Big)^2\sum_{j=0}\q^{2j}+
\q^4\t^2\Big((1+\q^2\t^2)(1+\q^4\t^2)+\q^6\t^4\Big)\Big(\sum_{i=0}\q^{2i}\Big)^2\sum_{j=0}(\q\t)^{2j}+
\q^{14}\t^6(1+2\t^2)\Big(\sum_{i,j=0}\q^{2(i+j)}\t^{2j}\Big)^2\Big]
+\nn\\
\!\!\!\!\!\!\!\!\!\!\!\!\!\!\!\!\!\!
+\a^4\q^4\t^6\Big[(1+\q^2)\Big((1+\q^2\t^2)(1+\q^4\t^2)+\q^{10}\t^4\Big)
+\q^8\t^4(1+\q^2\t^2+\q^4\t^4)\sum_{i=0}\q^{2i}+
\q^4\t^2(1+\q^6\t^4+\q^{12}\t^8)\sum_{i,j=0}\q^{2(i+j)}\t^{2i}\Big]\sum_{i,j=0}\q^{2(i+j)}\t^{2i}
+\ \ \ \nn\\
\left.+\a^6\q^{10}\t^{11}\Big[1+\q^2\t^2+\q^6\t^4\sum_{i=0}(\q\t)^{2i}\Big]
\Big[1+\q^2+\q^6\t^2\sum_{i=0}\q^{2i}\Big]
\sum_{i,j=0}\q^{2(i+j)}\t^{2i}
\right) \ \ \ \ \ \ \ \ \ \ \ \ \
\nn
\ee
}
%
Another positive series is given by the other four-point function with the same set of diagrams:
{\footnotesize
\be\label{HH3}
\widehat{\bf Z}_{cb}\left[\begin{array}{c|c}
[1]&[2]\cr
[2]&[1]\end{array}\right]\sim{(1+\a^2\t)(1+\a^2\t^3)(\q^2+\a^2\t)(1+\a^2\q^6\t^5)\over (1-\q^2)(1-\q^2\t^2)(1-\q^4\t^2)^2}
\left(\Big(1+\q^6\t^2\sum_{i=0}\q^{2i}\Big)\Big(1+\q^6\t^4\sum_{i=0}(\q\t)^{2i}\Big)+\right.
\nn\\\nn\\
\!\!\!\!\!\!\!
+\a^2\q^2\t\Big(1+\q^2\t^2+\t^4(1+\q^6\t^2+\q^8\t^2)\sum_{i=0}\q^{2i}+\q^6\t^4(1+\t^2+\q^2\t^2)
\sum_{i=0}(\q\t)^{2i}+
2\q^{12}\t^8\sum_{i=0}\q^{2i}\sum_{j=0}(\q\t)^{2j}\Big)+\nn\\
\left.+\a^4\q^4\t^6\Big(1+\q^2+\q^6\t^2\sum_{i=0}\q^{2i}\Big)
\Big(1+\q^2\t^2+\q^6\t^4\sum_{i=0}(\q\t)^{2i}\Big)\right) \nn
\ \ \ \ \ \ \
\ee
}

\noindent
As usual, when there are several positive expressions, one can wonder, if they are
somehow ordered, and look for a minimal one.
This is one more direction for a future study,
which, as we already mentioned, requires development of a more efficient technique
for complicated calculations with infinite series
than we have managed to find so far.



\end{document}